\documentclass[useAMS,usenatbib]{mn2e}
\usepackage{graphicx}
\usepackage{amssymb}\usepackage{hyperref}

\newcommand{\xmm}{{\it XMM-Newton} }

\title[Detection of radial velocity shifts due to black hole binaries near merger]{Detection of radial velocity shifts due to black hole binaries near merger}

\author[B. McKernan \& K.E.S. Ford]{B. McKernan$^{1,2,3,4}$\thanks{E-mail:bmckernan at amnh.org (BMcK)}, K.E.S. Ford$^{1,2,3,4}$ \\
$^{1}$Department of Science, Borough of Manhattan Community College, City University of New York, New York, NY 10007, USA\\
$^{2}$Department of Astrophysics, American Museum of Natural History, New York, NY 10024, USA\\
$^{3}$Graduate Center, City University of New York, 365 5th Avenue, New York, NY 10016, USA\\
$^{4}$Kavli Institute for Theoretical Physics, U. C. Santa Barbara, CA 93106, USA\\
}
\begin{document}

\date{Accepted. Received; in original form}

\pagerange{\pageref{firstpage}--\pageref{lastpage}} \pubyear{2013}

\maketitle

\label{firstpage}

\begin{abstract}
The barycenter of a  massive black hole binary will lie outside the event horizon of the primary black hole for modest values of mass ratio  and binary separation. Analagous to radial velocity shifts in stellar emission lines caused by the tug of planets, the radial velocity of the primary black hole around the barycenter can leave a tell-tale oscillation in the broad component of Fe~K$\alpha$ emission from accreting gas.  Near-future X-ray telescopes such as \emph{Astro-H} and \emph{Athena} will have the energy resolution ($\delta E/E \lesssim 10^{-3}$) to search nearby active galactic nuclei (AGN) for the presence of binaries with mass ratios $q \gtrsim 0.01$, separated by several hundred  gravitational radii. The general-relativistic and Lense-Thirring precession of the periapse of the secondary orbit imprints a detectable modulation on the oscillations. The lowest mass binaries in AGN will oscillate many times within typical X-ray exposures, leading to a broadening of the line wings and an over-estimate of black hole spin in these sources. Detection of periodic oscillations in the AGN line centroid energy will reveal a massive black hole binary close to merger and will provide an early warning of gravitational radiation emission. 
\end{abstract}


\section{Introduction}
\label{sec:intro}
Galactic nuclei host supermassive black holes ($>10^{6}M_{\odot}$) \citep{b4} and should 
be the sites of mergers of massive black holes (MBH). Binary black holes are expected to arise due to merging supermassive black holes (SMBH) throughout cosmic time in $\Lambda$CDM cosmology \citep[e.g.][]{springel05}, as well as due to intermediate mass black hole (IMBH) formation in disks around SMBH \cite[e.g.][]{lev07,imbh1}.
Possible observational signatures of these merging MBH are important to find \citep[e.g.][]{milo05,bog09,bode10,feka, imbh2,roedig14,farris15} as the search for binaries probes ever smaller separations \citep{comerford13, ford14,graham15}. Here we discuss the possibility of detecting those massive black hole binaries in the local Universe closest to merger, via a radial velocity imprint in the broad FeK$\alpha$ line. We demonstrate the effect and its magnitude and we show that near future X-ray telescopes with very fine energy resolution (\emph{Astro-H} and \emph{Athena} with $\delta E/E \lesssim 10^{-3}$ in the FeK band) will be able to detect MBH binaries near merger at mass ratios $q \gtrsim 0.01$ and separations of several hundred gravitational radii.

\section{Binary black hole dynamics}
\label{sec:binary}
\begin{figure}
\includegraphics[width=3.35in,height=3.35in,angle=-90]{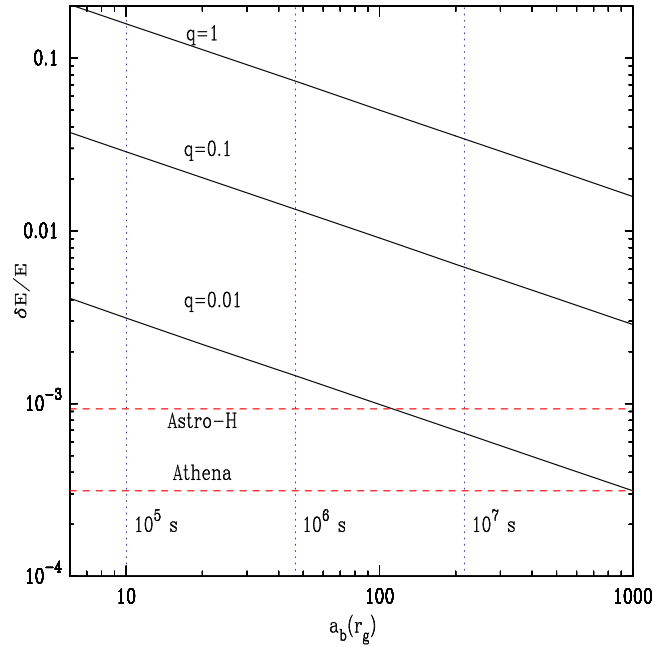}
\caption{The peak fractional change in energy $\delta E/E$ from eqn.~\ref{eq:dE} as a function of mass ratio $q$ and binary separation $a_{b}$. Horizontal dashed red lines denotes the planned energy resolution at 6.40keV for \emph{Astro-H}($\sim 6$eV) and \emph{Athena}(2eV). The blue vertical dotted lines mark the orbital separations correspond to oscillation periods around the barycenter of $10^{5,6,7}$s respectively, for a binary of mass $10^{8}M_{\odot}$. For comparison, \xmm EPIC has $\delta E/E \sim 0.025$, making it insensitive to most MBH binary parameter space.
\label{fig:mass_ratio}}
\end{figure}

A binary black hole will orbit its center of mass (barycenter), thereby adding a radial velocity component to line-emitting material immediately surrounding the primary SMBH. This is analagous to radial velocity shifts of stellar spectral lines as the star orbits its barycenter due to planetary tugs \cite[e.g.][]{ohta05}. By analogy, we can write the radial velocity component ($\rm{v}_{\rm rad}$) of the primary SMBH around the barycenter as \citep{ohta05}
\begin{equation}
\rm{v}_{\rm rad}=\frac{a_{\rm b}}{\ell}\frac{q}{1+q}\nu \  \rm{sin}(i)[\rm{sin} (f+\overline{\omega}) + \emph{e} \  \rm{sin}(\overline{\omega})]
\label{eq:vrad_ohta}
\end{equation}
where $q=m_{2}/M_{1}$ is the mass ratio of the secondary BH ($m_{2}$) to the primary BH ($M_{1}$), $a_{\rm b}$ is the binary semi-major axis, $\ell^{2}=1-e^{2}$ where $e$ is the orbital eccentricity of the secondary, $\nu=2\pi/T_{\rm orb}$ is the secondary mean motion or orbital frequency, $i$ is the inclination angle of the binary to the observers' sight line ($i=90^{\circ}$ is edge-on and yields maximal radial velocity components), $f$ is the true anomaly or the angle made by the secondary in its orbit as measured from a line joining the focus and periapse, and $\overline{\omega}$ is the argument of periapsis \footnote{Where the spin of the primary is important, $\Delta\overline{\omega}$ can be defined in terms of the precession of the orbital angular momentum of the secondary around the spin axis of the primary ($\Delta \Omega$) at fixed inclination and the precession of the periapse within the orbital plane ($\Delta \omega$), as $\Delta \overline{\omega}=\Delta \omega +\rm{cos} (i) \Delta \Omega$ \citep{merritt}.} (see e.g. Figs 1 \& 2 in \cite{ohta05} for sketches of these orbital parameters). The inner accretion disk is bound to the primary so spectral lines originating in the inner disk will oscillate around their rest centroid energies by $v_{\rm rad}/c$ on the timescale of the secondary's orbit \citep{yulu01,bog09,ses12}. As binary separation shrinks, the disk can become strongly perturbed. In the limit of $a_{b}$ of a few tens of $r_{g}$, we should expect the line profile to differ from the simple model below (see e.g. \citet{ses12,feka} and references in the latter for a discussion of the wide range of possible configurations including secondary disks, gaps and cavities). However most FeK$\alpha$ emission may originate within $\sim 20-30r_{g}$ of the primary and very tight binaries ($a_{b}$ of tens of $r_{g}$) will be much rarer than those at wider separations ($a_{b}$ of a few hundred $r_{g}$). Even if the line profile is perturbed, the periodic wobble will occur in the manner we outline below and may still be detectable. Eqn.~\ref{eq:vrad_ohta} can be rewritten as a energy shift ($\delta E$) in a spectral line centroid energy ($E$) as
\begin{equation}
\frac{\delta E}{E}=\left(\frac{r_{g}}{a_{\rm b}}\right)^{1/2}\frac{q}{1+q}\frac{\rm{sin}(i)}{\ell}[\rm{sin} (f+\overline{\omega}) + \emph{e}  \  \rm{sin}(\overline{\omega})]
\label{eq:dE}
\end{equation}
where $r_{g}=G(M_{1}+m_{2})/c^{2}$ is the gravitational radius of the binary and $a_{b}$ is in units of $r_{g}$. Fig.~\ref{fig:mass_ratio} shows the maximum $\delta E/E$ as a function of binary separation and mass ratio from eqn.~\ref{eq:dE} assuming circular orbits and edge-on inclination. Red horizontal dashed lines in Fig.~\ref{fig:mass_ratio} denote planned micro-calorimeter energy resolution for \emph{Astro-H} (upper; \citet{b29}) and the micro-calorimeter on \emph{Athena} (lower). The blue vertical dotted lines in Fig.~\ref{fig:mass_ratio} mark the orbital separations corresponding to periods of $10^{5}$s, $10^{6}$s and $10^{7}$s for a  $10^{8}M_{\odot}$ MBH binary. From Fig.~\ref{fig:mass_ratio}, radial velocity shifts in the broad  FeK$\alpha$ line due to MBH binaries with mass ratios $q\gtrsim 0.01$ and separations of several hundred $r_{g}$ could be measured and limits placed on the rate of hard MBH binary occurrence in the local Universe. Note that the secondary MBH is likely to have its own accretion disk. However, FeK$\alpha$ flux from the secondary in an strongly unequal mass binary is likely to be swamped by emission from the primary disk (luminosity $\propto M\dot{M}$), hence here we focus on emission from around the primary MBH.

\subsection{Precession of the secondary orbit}
\label{sec:precession}
As the MBH binary approaches merger, its orbit will tend to circularize due to emission of gravitational radiation, however other factors may act against this effect (see below and e.g. \cite{cuadra09,roedig12}). As the binary orbit shrinks below $a_{b} \sim 10^{3-4}r_{g}$, general relativistic (GR) effects cause orbital changes that may be detectable in oscillations of the centroid energy of the broad FeK$\alpha$ line. Changes in $\overline{\omega}$ over time will have the effect of modulating the sinusoidal function of $\delta E/E$ in eqn.~\ref{eq:dE}. 
The overall precession angle evolves as
\begin{equation}
\overline{\omega}(t)=\overline{\omega}(0) +[-\dot{\overline{\omega}}_{\rm disk} + \dot{\overline{\omega}}_{LK}+ \dot{\overline{\omega}}_{GR} + \dot{\overline{\omega}}_{LT} +\dot{\overline{\omega}}_{Q}]t
\label{eq:evolve_precession}
\end{equation}
where $\overline{\omega}(0)$ is the initial angle value, $\dot{\overline{\omega}}_{\rm disk}$ is the rate of  (retrograde) precession of periapse due to the interaction between the secondary and the disk (analagous to the gravitational tugs between planets in the planetary case), $\dot{\overline{\omega}}_{LK}$ is the rate of precession of periapse due to the interchange of (e,i) due to the Lidov-Kozai (LK) mechanism, $\dot{\overline{\omega}}_{GR}$ is the (geodetic) rate of precession of periapse due to a Schwarzschild spacetime, $\dot{\overline{\omega}}_{LT}$ is the precession rate of periapse due to Lense-Thirring (LT) frame-dragging by a  spinning primary and $\dot{\overline{\omega}}_{Q}$ is the rate of  precession of periapse due to the quadrupole moment of the non-spherical (spinning) primary. Note that LT frame-dragging effects can lead to inner disk warping \citep{bp75}, so the angle of the inner disk to the observers sightline can be independent of the angle of orbital plane of the secondary ($i$ in eqn.\ref{eq:dE}). 

In planetary studies of the effect of planetary tugs on the radial velocity component of stars, $\overline{\omega}$ tends to evolve slowly with time and the effect of co-orbital planets (equivalent to our $\dot{\overline{\omega}}_{\rm disk}$ term above), dominates over other terms. However, in the case of a merging MBH binary, GR effects become important and we can approximate the precession of periapse of the secondary MBH orbit per $2\pi$ revolution as \citep{merritt}
\begin{equation}
\delta \overline{\omega} \approx 6\pi F \left[1 \pm \frac{4s^{2}}{3}F^{1/2} \rm{cos}(i)+ \frac{s^{2}}{4}F(1-4\rm{cos}^{2}(i))\right] + \delta{\overline{\omega}}_{\rm disk} 
\label{eq:dwapprox}
\end{equation}
where the term in square brackets is $[\delta{\overline{\omega}}_{\rm GR}+\delta{\overline{\omega}}_{\rm LT}+\delta{\overline{\omega}}_{\rm Q}]$,  $F=r_{g}/(a_{b}\ell^{2})$, $s=\pm[0,1]$ is the dimensionless spin parameter of the primary MBH and $\delta{\overline{\omega}}_{\rm disk}=2\pi q_{a_{b}} \ell/(1+\ell)$, where  $q_{a_{b}}=M_{(r<a_{b})}/M_{1}$ is the mass ratio of distributed (disk) mass at $r<a_{b}$ to mass $M_{1}$. The  $\delta{\overline{\omega}}_{GR}=6\pi F$ term dominates $\delta{\overline{\omega}}_{\rm disk}$ when $\ell^{2} \leq (3 r_{g}/a_{b}q_{a_{\rm b}})^{2/3}$, which holds if $a_{b}/r_{g}<M_{1}/M_{(r<a_{b})}$ for modest $e$, or broadly if $a_{b} \leq 10^{3-4}r_{g}$ for most realistic disk models. The contribution of spin terms ($\delta{\overline{\omega}}_{\rm spin}=\delta{\overline{\omega}}_{\rm LT}+\delta{\overline{\omega}}_{\rm Q}$) to the precession rate is small at wide binary separations, growing from $\delta{\overline{\omega}}_{\rm spin}/\delta{\overline{\omega}}_{\rm GR} \sim 0.02$ at $10^{3}r_{g}$ to $\delta{\overline{\omega}}_{\rm spin}/\delta{\overline{\omega}}_{\rm GR} \sim 0.2$ at 10$r_{g}$.  We ignore the effect of the LK mechanism ($\delta{\overline{\omega}}_{\rm LK}$) since it requires a tertiary (disk or satellite) at high relative inclination and more massive than the secondary \citep{ford00}. However, since stalled SMBH are expected to occur on $\sim$pc scales in galactic nuclei \citep{milo01}, it is possible that the LK mechanism could yield binary eccentricities $e>0$ in randomly observed MBH binaries even at very small ($\sim 100r_{g}$) separation.

\begin{figure}
\includegraphics[width=3.35in,height=3.35in,angle=-90]{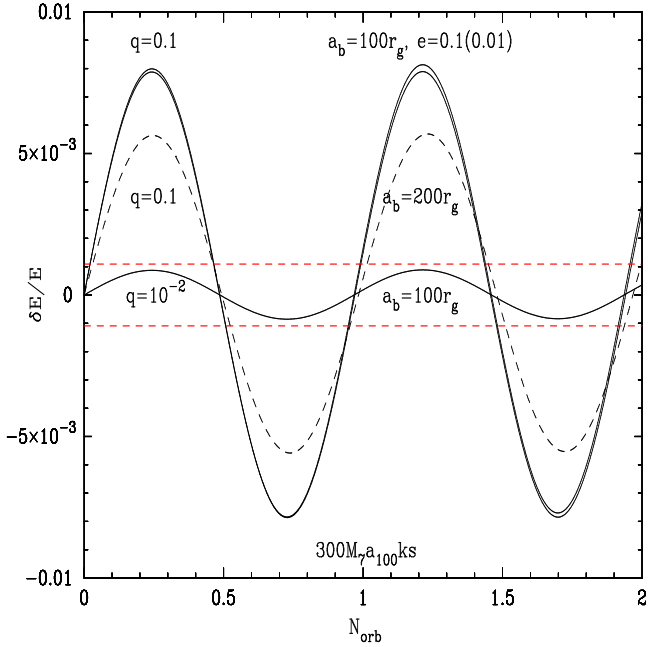}
\caption{$\delta E/E$ over two periods of oscillation of the barycenter of a MBH binary inclined at $60^{\circ}$ to the observers' sightline ($90^{\circ}$ is edge-on). Red dashed lines represent the expected limits of $\delta E/E$ for \emph{Astro-H}. A $300M_{7}a_{100}$ks exposure corresponds to one orbit at $100r_{g}$ around a $M_{7}=M_{1}/10^{7}M_{\odot}$ SMBH.    
\label{fig:2orbits}}
\end{figure}

Fig.~\ref{fig:2orbits} shows $\delta E/E$ for the centroid energy of the broad FeK$\alpha$ line as a function of a $2\pi$ orbit over two periods of oscillation of the primary about the binary barycenter. Solid lines depict MBH binaries (with $e=0.01,0.1$) at $a_{b}=100r_{g}$ and the black dashed line depicts an MBH binary at $a_{b}=200r_{g}$. A $q=0.01$ binary at $a_{b}=100r_{g}$ yields an oscillation that is not detectable with \emph{Astro-H} (dashed red lines) but is detectable with \emph{Athena} ($\delta E/E \sim 3\times 10^{-4}$). Even over only two periods of oscillation, the effects of GR precession on different binary configurations are starting to appear in the slight phase shift between the $e=0.01,0.1$ curves and between the $a_{b}=100,200r_{g}$ curves in Fig.~\ref{fig:2orbits}.

\begin{figure}
\includegraphics[width=3.35in,height=3.35in,angle=-90]{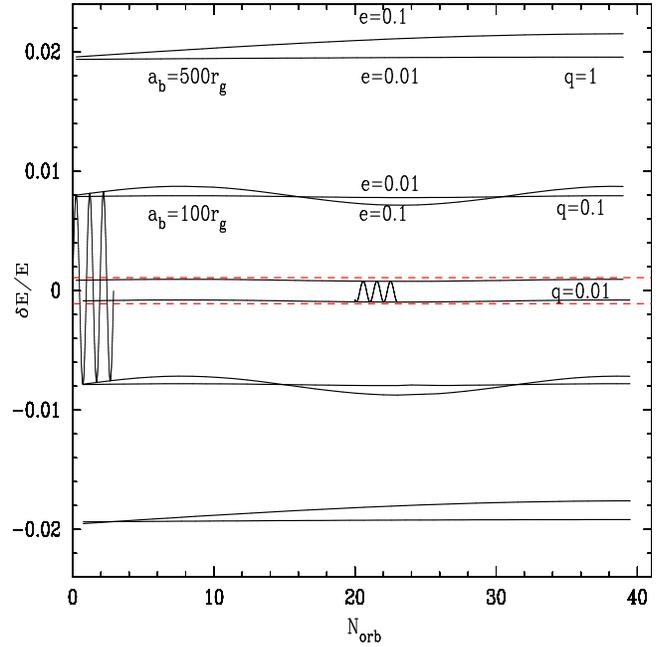}
\caption{As Fig.~\ref{fig:2orbits} except over 40 oscillation of the barycenter of an MBH binary. Curves  with $|\delta E/E| \sim 0.02$ show the envelope of oscillations for $q=1$, $a_{b}=500r_{g}$ at eccentricities $e=0.1$(upper) and $e=0.01$(lower). Curves with $|\delta E/E| \sim 0.008$ show the envelope of oscillations (first three are depicted) for $q=0.1, a_{b}=100r_{g}$ at eccentricities $e=0.01,0.1$. 
Curves with $|\delta E/E| \sim 0.001$ show the envelope of oscillations (oscillations 20-23 are shown) for $q=0.01, a_{b}=100r_{g}$ at eccentricity $e=0.1$. Red dashed lines represent the expected limits of $\delta E/E$ for \emph{Astro-H}. Precession of periapse (by $\sim 11^{\circ}$/orbit at $a_{b}=100r_{g}$) leads to periodicity in $\delta E/E$ after $\sim 33$ orbits. A $1.2$Ms exposure corresponds to $40$ orbits at $100r_{g}$ around a $M_{6}=M_{1}/10^{6}M_{\odot}$ SMBH.   
\label{fig:40orbits}}
\end{figure}

Fig.~\ref{fig:40orbits} shows an expanded version of Fig.~\ref{fig:2orbits}, showing $\delta E/E$ for the centroid energy of the broad FeK$\alpha$ line as a function of $2\pi$ orbit, over $40$ orbits of the primary about the binary barycenter. The two periods of oscillation from Fig.~\ref{fig:2orbits} for $q=0.1,a_{b}=100r_{g}$ are shown in Fig.~\ref{fig:40orbits} as a part of a three period oscillation between $\pm0.008$. The envelope for these oscillations assuming $e=0.01,0.1$ is extended to $40$ orbits, revealing the modulating effects of GR-dominated orbital precession of the secondary, showing up in oscillations of the primary around the barycenter.  Also shown in Fig.~\ref{fig:40orbits} are the envelope of oscillations of the barycenter of a $q=0.01,a_{b}=100r_{g}, e=0.1$ MBH binary together with oscillations $20-23$ which will therefore not be detectable with \emph{Astro-H} (red dashed lines), but will be detectable with \emph{Athena} (assuming $\delta E/E \sim 3 \times 10^{-4}$). Fig.~\ref{fig:40orbits} also shows the limiting case of the envelope of oscillations due to a $q=1$ MBH binary at $a_{b}=500r_{g}$ with $e=0.1,0.01$ (assuming only one MBH has an associated broad FeK$\alpha$ line for simplicity). Note that a $1.2M_{6}$Ms exposure corresponds to $40$ orbits at $100r_{g}$ around a $M_{6}=M_{1}/10^{6}M_{\odot}$ SMBH (see observing strategies below).

\subsection{The broad and narrow FeK$\alpha$ line components}
\label{sec:broad}
Broad FeK$\alpha$ lines are observed in several nearby active galactic nuclei (AGN) \citep[e.g.][]{nan97,braito07} and originate in fluorescent Fe deep in the gravitational potential well of the central supermassive black hole \citep[see e.g.][for a review]{reynow03}. See \citet{feka} (and references therein) for a discussion of how the broad FeK$\alpha$ line component can be used to probe configurations of gas in the inner accretion disk and therefore test for possible MBH binaries. The narrow core of the FeK$\alpha$ line observed in AGN \citep[e.g.][]{yp04} originates in fluorescing cold gas (Fe~{\sc i}-Fe~{\sc xvii}) far from the supermassive black hole \citep{shu10}. We expect variations of the narrow core of the FeK$\alpha$ line to be uncorrelated on observing timescales with variations in the broad component and, more importantly, the narrow line centroid should not vary. Therefore, without knowing the absolute value of $\delta E/E$ in the broad line in a given AGN, we can monitor its variation over time, by comparing the centroid of the narrow line with the oscillating centroid of the broad component.

\begin{figure}
\includegraphics[width=3.35in,height=3.35in,angle=-90]{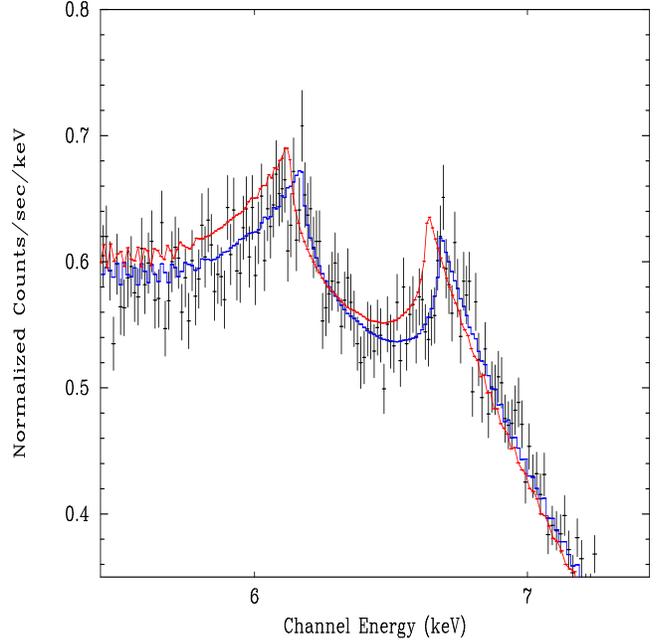}
\caption{A simulated 100ks observation with the \emph{Athena} microcalorimeter of an AGN at $z=0.01$, with a 2-10keV flux of $4.5 \times 10^{-12} \rm{erg} \rm {cm}^{-2} \rm{s}^{-1} $, which is generated by a diskline model \citep{fab89} (see text for details). Black data points correspond to simulated data where the average line centroid shift is $\delta E/E=+0.004$, which is the average of the first positive oscillation in Fig.~\ref{fig:2orbits}) and the blue solid line corresponds to the best-fit model to these data. The red solid line corresponds to the best model fit when $\delta E/E=-0.004$, or the average $\delta E/E$ for the first negative oscillation in Fig.~\ref{fig:2orbits} (simulated data not shown for clarity). The 200ks oscillation period corresponds to a $q=0.1$ binary at $a_{b}=100r_{g}$ with $M_{1}=6 \times 10^{6}M_{\odot}$.
\label{fig:sim_ixo}}
\end{figure}

Fig.~\ref{fig:sim_ixo} shows a $100$ks simulated\footnote{http://heasarc.gsfc.nasa.gov/cgi-bin/webspec} observation with the planned micro-calorimeter on \emph{Athena} of an AGN at $z=0.01$ with a 2-10keV photon flux of $4.5 \times 10^{-12} \rm{erg}  \rm{cm^{-2}} \rm{s^{-1}}$. The simulated data shown (binned at $\sim 12$eV) correspond to a $100$ks exposure with $\delta E/E$ equal to the average from the first positive oscillation for $q=0.1, a_{b}=100r_{g}$ in Fig.~\ref{fig:2orbits} (or $\delta E/E=4 \times 10^{-3}$. The blue curve corresponds to the best-fit model to these data and the red curve shows the best-fit model to an equivalent $100$ks observation with $\delta E/E=-0.004$, equivalent to the average $\delta E/E$ for orbits 0.5-1.0 in Fig.~\ref{fig:2orbits}. The displacement between the two model fits by eye highlights how the oscillation of binaries around their barycenter will be easily detectable with \emph{Athena}. 
The narrow FeK$\alpha$ line component (ignored in this model) should maintain the same centroid energy, allowing the relative shift in $\delta E/E$ to be determined. The \textsc{XSPEC} toy model for the data in Fig.~\ref{fig:sim_ixo} was \textsc{zphabs(zpowerlw+diskline)} where the absorbing Hydrogen column was $10^{21}\rm{cm}^{-2}$ and the powerlaw index was $-1.8$. The primary diskline for the simulated data had an inclination angle of $60^{\circ}$ and emissivity index $k=-2.5$, as well as a line normalization of $0.1$ with respect to the continuum to illustrate the effect of the radial-velocity shift. 

\section{Observing strategies}
\label{sec:strategies}
From Fig.~\ref{fig:2orbits}, a $2\pi$ barycenter oscillation period for a $q=0.1, a_{b}=100r_{g}$ binary will take $\sim 1$year around a $M_{1}=10^{9}M_{\odot}$ primary SMBH, but only $\sim 8$ hours around a $M_{1}=10^{6}M_{\odot}$ primary SMBH. Since typical exposures of hundreds of ks (few days) are required to build up line fluxes with modern X-ray telescopes, techniques to search for MBH binaries in FeK$\alpha$ lines from AGN must depend on the moderately accurate estimates of the central mass \citep[e.g.][]{gult09,mck10,b4}. Large mass ($\sim 10^{9}M_{\odot}$) SMBH are the simplest to test for barycenter oscillations, since a MBH binary will yield a monotonic change in the FeK$\alpha$ centroid energy between exposures over year ($a_{b} \sim 100r_{g}$) to multi-year ($a_{b} >100r_{g}$) timescales. Moderate mass ($\sim 10^{7}-10^{8}M_{\odot}$) SMBH will yield an oscillation over a single long exposure-- a $300M_{7}a_{100}$ks exposure corresponds to one orbit at $a_{b}=100r_{g}$ around a $M_{7}=M_{1}/10^{7}M_{\odot}$ SMBH. In these cases, a division of the exposures into two or three segments, at different phases, is sufficient to test for MBH binaries at moderate separations.
The smallest mass ($\sim 10^{6}M_{\odot}$) SMBH will be the most difficult to test for MBH binaries, since a several hundred ks exposure might contain tens of oscillations. The oscillations will broaden the wings of the broad FeK$\alpha$ component and models measuring the spin of a single SMBH \citep[e.g.][]{bren11} will therefore over-estimate of the spin of the SMBH. To test for tight MBH binaries in low mass AGN, and constrain $q,a_{b}$ parameter space, long exposures must be chopped into periodic odd-even sets on short timescales at different phases, that may be co-added to test for radial velocity oscillations and modulation by precession effects. Note that the effective area of \emph{Athena} (which falls off at higher energies) may be well exploited by AGN at modest redshifts.  

\section{Conclusions}
\label{sec:conclusions}
Analagous to the search for planets around stars using radial velocity shifts in stellar spectral lines, radial velocity shifts will be imprinted on the broad FeK$\alpha$ line due to oscillations of a massive black hole binary (MBHB) around its barycenter. Near future X-ray telescopes (\emph{Astro-H} or \emph{Athena}) will have the energy resolution to search for such oscillations in the broad FeK$\alpha$ line observed in active galactic nuclei. We demonstrate that oscillations of MBHB about their barycenter with mass ratios $ \gtrsim 0.01$, moderate binary separations ($a_{b} \lesssim 1000r_{g}$) and moderate eccentricities ($e \sim 0.01-0.1$) will be detectable in multiple exposures of the broad FeK$\alpha$ line over timescales appropriate for the MBH mass. The periodic red- or blue-shift of the broad line energy centroid can be calibrated against the centroid of the narrow FeK$\alpha$ line.  The magnitude of precession of the secondary MBH orbit is dominated by GR effects and we demonstrate that this will be detectable for modest mass ratios and separations. Following Bayesian searches for radial velocity effects in stars due to tugs by unobserved planets\citep[e.g.][]{brewer15}, we can take a similar approach to finding periodic signal in integrated (and multi-epoch) exposures of broad FeK$\alpha$ lines in AGN. Observations of such oscillations will reveal the closest MBH binaries and brightest gravitational wave sources in the local Universe.

\section*{Acknowledgements}
We acknowledge support from grants NSF PAARE AST-1153335 and NSF PHY 11-25915 and useful discussions with Tamara Bogdanovic, Andy Fabian, Zoltan Haiman and Daniel Stern. We thank Allyn Tennant for maintaining Web-QDP and HEASARC at NASA GSFC for maintaining Webspec.


\label{lastpage}

\end{document}